\documentclass[12pt]{article}
\pdfoutput =1
\usepackage{float} 
\usepackage{amsmath}
\usepackage{slashed}
\usepackage{amsfonts}   
\usepackage{amssymb}

\begin{document}
\date{}
\title{{\bf{\Large Nonrelativistic giant magnons from Newton Cartan strings}}}
\author{
 {\bf {\normalsize Dibakar Roychowdhury}$
$\thanks{E-mail:  dibakarphys@gmail.com, dibakarfph@iitr.ac.in}}\\
 {\normalsize  Department of Physics, Indian Institute of Technology Roorkee,}\\
  {\normalsize Roorkee 247667, Uttarakhand, India}
\\[0.3cm]
}

\maketitle
\begin{abstract}
We show nonrelativistic (NR) giant magnon dispersion relations by probing the torsional Newton Cartan (TNC) geometry with (semi)classical nonrelativistic rigidly rotating strings. We construct NR sigma models over $ R \times S^2 $ and consider two specific limiting cases those are of particular interest. Both of these limiting conditions give rise to what we identify as the \emph{small} momentum limit of the giant magnon dispersion relation in the dual SMT at \emph{strong} coupling. We further generalize our results in the presence of background NS-NS fluxes. Our analysis reveals that unlike its relativistic counterpart, the NR string theory lacks of single spike solutions.
\end{abstract}
\section{Overview and Motivation}
The quest for a UV complete theory of nonrelativistic (NR) gravity gives rise to what is known as the NR formulation of string theory \cite{Gomis:2000bd}-\cite{Gomis:2005pg} over curved manifolds. Recent advances along this line of research reveals the existence of two parallel formulations of NR string sigma models those at the first place appear to be quite distinct from each other. One of these formulations goes under the name of Newton-Cartan (NC) formulation of NR string sigma models \cite{Andringa:2012uz}-\cite{Yan:2019xsf}  whereas on the other hand, the other approach is based on the so called \emph{null} reduction of Lorenzian manifolds known as toraional Newton-Cartan (TNC) geometry\footnote{Recently it has been argued that both of these approaches are in fact equivalent to each other \cite{Harmark:2019upf}. However, it is the second approach that is of particular interest as far as the present analysis is concerned.}\cite{Harmark:2017rpg}-\cite{Roychowdhury:2019sfo}. It has been conjectured that $ 1/c $ limit of TNC sigma model (thus obtained via null reduction of $ AdS_5 \times S^5 $ (super)strings) is dual to $ \lambda \rightarrow 0 $ limit of the full $ \mathcal{N}=4 $ SYM spectrum on $ R\times S^3 $ known as the Spin Matrix Theory (SMT) \cite{Harmark:2017rpg},\cite{Harmark:2014mpa}. Keeping the spirit of the above conjectured duality, the purpose of the present article is therefore to check whether the NR analogue of the relativistic giant magnon spectrum \cite{Beisert:2005tm}-\cite{Lee:2008sk} is actually reproducible considering $ 1/c $ limit of TNC sigma models on $ R\times S^2 $. Below we further elaborate on this taking specific example from $ \mathcal{N}=4 $ SYM. 

The single spin magnon excitation in $ \mathcal{N}=4 $ SYM theory is expressed in the form of the dispersion relation \cite{Beisert:2005tm},
\begin{eqnarray}
E -J =\sqrt{1+\frac{\lambda}{\pi^2}\sin^2 \frac{p_m}{2}}
\label{ee1}
\end{eqnarray}
where, $ \lambda $ is the coupling constant in the dual gauge (SYM) theory. On the other hand, the effective t'Hooft coupling ($ \mathtt{g} $) of the SMT \cite{Harmark:2014mpa} is related to the $ \mathcal{N}=4 $ SYM coupling with the rescaling of the form, $  \mathtt{g}= c^2 \lambda $ where $ c(\rightarrow \infty) $ is the speed of light and $ |\lambda| \ll 1 $ such that $  \mathtt{g} $ is \emph{finite}. Rewriting (\ref{ee1}) in terms of SMT coupling and considering the \emph{small} momentum ($ |p_m| \ll 1 $) limit we find,
\begin{eqnarray}
E -J \sim \sqrt{1+\frac{\mathtt{g}}{4\pi^2}\Big|\frac{p_m}{c}\Big|^{2}}
\end{eqnarray} 
which in the limit of \emph{strong} ($ \mathtt{g}\gg1 $) coupling finally yields,
\begin{eqnarray}
E -J \sim \frac{\sqrt{\mathtt{g}}}{2 \pi}\Big|\frac{p_m}{c}\Big| 
\label{magnon}
\end{eqnarray}
where we understand the limit, $ \Big|\frac{p_m}{c}\Big| \ll 1 $ such that the R.H.S. of (\ref{magnon}) is \emph{finite} in the limit of strong coupling. The purpose of the present paper is therefore to reestablish the above strong coupling result (\ref{magnon}) from \emph{classical} NR sigma model calculations on $ R \times S^2 $ \cite{Roychowdhury:2019olt} where we identify the giant magnon momentum on the R.H.S. of (\ref{magnon}) as being some \emph{effective} geometrical entity ($ | \Delta \varphi_{m}| $) on the dual string theory side \cite{Hofman:2006xt} at strong coupling. 

As a further continuation of our analysis, we explore single spin giant magnon dispersion relation in the presence of background NS-NS fluxes. There have been several reasons for studying giant magnon as well as single spike solutions in the presence of NS-NS fluxes \cite{Huang:2006vz}-\cite{Lee:2008sk}. The most important one goes under the name of ``flux stabilization" which states that the background NS-NS fluxes sort of stabilizes D branes (wrapping $ S^2 $) from shrinking it to zero size. The Page charge associated with such D brane configurations is measured in terms of flux associated with the two sphere namely, $\mathcal{Q}_{D} \sim \int_{S^2}\mathcal{B} $ which takes quantized values on the world-volume of the D brane. The other reason for considering NS-NS fluxes stems from the fact that the background magnetic filed naturally breaks supersymmetry as it couples differently with particles (magnons) of different spins \cite{Huang:2006vz}. Therefore such a configuration is more appropriate in a less supersymmetric setup which is what we are concerned here.
\section{Nonrelativistic magnons}
\subsection{NR sigma model}
We start with the Nambu-Goto (NG) action for TNC strings \cite{Harmark:2017rpg} on $ R \times S^2 $,
\begin{eqnarray}
\mathcal{S}_{NG} = \frac{\sqrt{\lambda}}{2\pi}\int d\tau d\sigma \mathcal{L}_{NG}~;~\sqrt{\lambda}=\frac{L^2}{\alpha'}
\label{e1}
\end{eqnarray}
where, the corresponding Lagrangian density could be formally expressed as \cite{Roychowdhury:2019olt},
\begin{eqnarray}
\mathcal{L}_{NG}=\frac{\varepsilon^{\alpha \alpha'}\varepsilon^{\beta \beta'}}{\varepsilon^{\alpha \alpha'}\chi_{\alpha}\partial_{\alpha'}\zeta}(\partial_{\alpha'}\zeta \partial_{\beta'}\zeta -\chi_{\alpha'}\chi_{\beta'})\left( \partial_{\alpha}\theta \partial_{\beta}\theta +\sin^{2}\theta\partial_{\alpha}\varphi \partial_{\beta}\varphi\right)-\varepsilon^{\alpha \beta}\cos\theta \partial_{\alpha}\varphi \partial_{\beta}\zeta.
\label{e2}
\end{eqnarray}
Here, $ \zeta $ is the additional \emph{compact} direction associated with the null reduced target space geometry along which the string winds \cite{Harmark:2017rpg}-\cite{Harmark:2018cdl}. On the other hand, $ \varepsilon^{01}=-\varepsilon_{01}=1 $ is the 2D Levi-Civita symbol together with, 
\begin{eqnarray}
\chi_{\alpha}=2 \partial_{\alpha}t +\partial_{\alpha}\psi -\cos\theta \partial_{\alpha}\varphi.
\end{eqnarray}

The next step would be to take the so called large $ c (\rightarrow \infty)$ limit \cite{Harmark:2017rpg} associated with the world-sheet d.o.f. which thereby yields the NR sigma model,
\begin{eqnarray}
\tilde{\mathcal{S}}_{NG}=\frac{\sqrt{\mathtt{g}}}{2\pi} \int d\tau d\sigma \tilde{\mathcal{L}}_{NG}
\end{eqnarray}
together with the rescaled Lagrangian (density) of the following form \cite{Roychowdhury:2019olt},
\begin{eqnarray}
\tilde{\mathcal{L}}_{NG}= \frac{\varepsilon^{\alpha\alpha'}\varepsilon^{\beta\beta'}}{\partial_{\tau}t}\partial_{\alpha'}t\partial_{\beta'}t (\partial_{\alpha}\theta \partial_{\beta}\theta +\sin^2\theta \partial_{\alpha}\varphi \partial_{\beta}\varphi)+\varepsilon^{\alpha \beta}\cos\theta \partial_{\alpha}\varphi \partial_{\beta}\zeta  +\mathcal{O}(c^{-2})
\label{e4}
\end{eqnarray}  
where, $ \mathtt{g}=c^2 \lambda $ is the \emph{effective} string tension \cite{Harmark:2018cdl} in the NR limit. Following NR sigma model/SMT duality \cite{Harmark:2014mpa} we understand $ \mathtt{g}(\gg 1) $ as the coupling constant on the dual gauge theory side such that the corresponding SYM coupling $ (\lambda) $ is considered to be \emph{weak}.

In order to proceed further, we choose the following parametrization \cite{Ishizeki:2007we} for the rigidly rotating NR strings over $ R \times S^2 $ namely,
\begin{eqnarray}
t = \kappa \tau ~;~\varphi = \varpi \tau + \sigma ~;~\theta =\theta (\sigma)~;~\zeta = \sigma 
\label{e5}
\end{eqnarray}
which upon substitution into (\ref{e4}) yields,
\begin{eqnarray}
\tilde{\mathcal{L}}_{NG}=\kappa (\theta'^2 +\sin^2\theta)+\varpi \cos\theta +\mathcal{O}(c^{-2}).
\label{e6}
\end{eqnarray}

The equation of motion that readily follows from (\ref{e6}) could be formally expressed as,
\begin{eqnarray}
\theta'' -\sin\theta \left(\cos\theta -\frac{\varpi}{2 \kappa} \right) =0.
\label{e7}
\end{eqnarray}

The above equation (\ref{e7}) could be integrated once to obtain,
\begin{eqnarray}
\theta'(\sigma)= \sqrt{(\xi^{2}-\eta^2 \sin^{2}\frac{\theta}{2})(\beta^{2}\sin^2\frac{\theta}{2}-\alpha^2)}
\label{e8}
\end{eqnarray}
which is subjected to the realization of the following constraints, 
\begin{eqnarray}
\alpha^2 \xi^2 &=& \frac{1}{2}-\frac{\mathcal{C}+\varpi}{\kappa}\\
\eta^2 \alpha^2 +\xi^2 \beta^2 &=& 4 \left( 1-\frac{\varpi}{2\kappa}\right) \\
\eta^2 \beta^2 &=&4
\end{eqnarray} 
where $ \mathcal{C} $ being the constant of integration. In the subsequent analysis, we further impose constraints on the parameter space of solutions namely, $ \frac{\varpi}{\kappa}\ll 1 $ and $  \frac{\mathcal{C}+\varpi}{\kappa}\ll 1$. 
\subsection{The spectrum}
Our next step would be to compute the conserved charges associated with the 2D sigma model. We first note down the energy associated with the NR stringy configuration,
\begin{eqnarray}
\mathcal{E}= \frac{\sqrt{\mathtt{g}}}{2\pi}\int\frac{(\xi^{2}-\eta^2 \sin^{2}\frac{\theta}{2})(\beta^{2}\sin^2\frac{\theta}{2}-\alpha^2)+\sin^2\theta}{\sqrt{(\xi^{2}-\eta^2 \sin^{2}\frac{\theta}{2})(\beta^{2}\sin^2\frac{\theta}{2}-\alpha^2)}}d\theta.
\label{e12}
\end{eqnarray}  

Like in the relativistic example \cite{Ishizeki:2007we}, requiring the fact that the above entity (\ref{e12}) to be real, we find the following two bounds on the azimuthal angle namely, (I) $ \frac{\alpha^2}{\beta^2}<\sin^2\frac{\theta}{2}<\frac{\xi^2}{\eta^2} $ and (II) $ \frac{\xi^2}{\eta^2} <\sin^2\frac{\theta}{2}<\frac{\alpha^2}{\beta^2}$. Keeping these facts in mind, our next step would be to explore the dispersion relations in the following two limits namely, $ |\eta| \rightarrow \xi $ as well as $ | \beta| \rightarrow \alpha $. 

However, before getting into that, it is customary first to note down the second conserved quantity namely the angular momentum,
\begin{eqnarray}
\mathcal{J}_{\varphi}=\frac{\sqrt{\mathtt{g}}}{2\pi}\int\frac{\cos\theta}{\sqrt{(\xi^{2}-\eta^2 \sin^{2}\frac{\theta}{2})(\beta^{2}\sin^2\frac{\theta}{2}-\alpha^2)}}d\theta
\label{E17}
\end{eqnarray}
as well as the angular difference between the end points of the solitonic excitation,
\begin{eqnarray}
\Delta \varphi =\int\frac{d\theta}{\sqrt{(\xi^{2}-\eta^2 \sin^{2}\frac{\theta}{2})(\beta^{2}\sin^2\frac{\theta}{2}-\alpha^2)}}.
\label{E18}
\end{eqnarray}
With these above set up in hand, we are now in a position to explore various limiting conditions associated with the 2D sigma model.

The first limiting case we are interested in is $ |\eta| \rightarrow \xi $. In order to explore this limit, it is first customary to rewrite (\ref{e12}) as,
\begin{eqnarray}
\mathcal{E}= \frac{\sqrt{\mathtt{g}}}{2\pi}\int_{\gamma_{min}}^{\gamma_{max}}\frac{4(\sin^{2}\gamma_{max}- \sin^{2}\frac{\theta}{2})(\sin^2\frac{\theta}{2}-\sin^{2}\gamma_{min})+\sin^2\theta}{2\sqrt{(\sin^{2}\gamma_{max}- \sin^{2}\frac{\theta}{2})(\sin^2\frac{\theta}{2}-\sin^{2}\gamma_{min})}}d\theta.
\label{E19}
\end{eqnarray}
where we set the upper limit of the integral\footnote{The physical picture that we have in mind is that of a string soliton wrapping along the equator of $ S^2 $ whose excitations (magnons) travel with a specific momentum, $ |p_m |\ll 1 $.},
\begin{eqnarray}
\gamma_{max}=\sin^{-1}\Big| \frac{\xi}{\eta}\Big| = \frac{\theta_{max}}{2}
\end{eqnarray}
together with the lower limit,
\begin{eqnarray}
\gamma_{min}=\sin^{-1}\Big| \frac{\alpha}{\beta}\Big| = \frac{\theta_{min}}{2}.
\end{eqnarray}
Clearly, the limit $ |\eta| \rightarrow \xi $ stands for setting $\gamma_{max}=\pi/2 $. 

Next, we implement the above limit into (\ref{e8}) and find,
\begin{eqnarray}
\sigma (\theta)= \pm \frac{1}{2}\int\frac{d\theta}{\cos\frac{\theta}{2}\sqrt{\sin^2\frac{\theta}{2}-\sin^{2}\gamma_{min}}}=\pm\frac{ \tanh ^{-1}\left(\frac{\sqrt{2}\sin \frac{\theta }{2} \cos  \gamma _{\min }}{\sqrt{|\cos \left(2 \gamma _{\min }\right)-\cos \theta}|}\right)}{\cos \gamma _{\min}}.
\label{e18}
\end{eqnarray}

Using (\ref{e18}), the angular difference between the end points of the string soliton turns out to be\footnote{Here we rescale the azimuthal angle, $ |\Delta \varphi_{m} |\equiv \cos\gamma_{min}|\Delta \varphi|$ which we further identify as being the \emph{effective} geometrical angle which corresponds to giant magnon momentum ($ |p_m| \ll 1 $) in the dual SMT sector at strong coupling.},
\begin{eqnarray}
 |\Delta \varphi_{m} |\simeq\coth ^{-1}\left(\sqrt{|1-\tan ^2\gamma _{\min }|}\right)-\tanh ^{-1}\left(\frac{\sqrt{2}\sin\frac{\gamma _{\min }}{2} \cos\gamma_{min}}{\sqrt{| \cos \left(2 \gamma _{\min }\right)-\cos \gamma _{\min }|}}\right).
\label{e19}
\end{eqnarray}
which for a specific value $ \gamma_{min}=\frac{\pi}{3} $ yields,
\begin{eqnarray}
|\Delta \varphi_{m} |\sim \coth ^{-1}\left(\frac{3}{\sqrt{2}}\right).
\end{eqnarray}

On the other hand, a straightforward computation reveals the dispersion relation,
\begin{eqnarray}
\mathcal{E} -\mathcal{J}_{\varphi}= \frac{\sqrt{\mathtt{g}}}{2\pi}\left( |\Delta \varphi |+2 \Delta\jmath \right) 
\label{e24}
\end{eqnarray}
where we note down the constant shift,
\begin{eqnarray}
\Delta\jmath = \log \left(\sqrt{2} \sin\frac{\gamma _{\min }}{2}+\sqrt{|\cos \left(2 \gamma _{\min }\right)-\cos \gamma _{\min }}|\right)-\log \left(1+\sqrt{|\cos \left(2 \gamma _{\min }\right)|}\right)\nonumber\\
+\sqrt{|\cos \left(2 \gamma _{\min }\right)|}-\sqrt{2} \sin\frac{\gamma _{\min }}{2} \sqrt{|\cos \left(2 \gamma _{\min }\right)-\cos\gamma _{\min }|}.
\label{e25}
\end{eqnarray}

In order to identify (\ref{e24}) to that with the actual magnon dispersion relation (\ref{magnon}) below we define the difference between the limits of the integral,
\begin{eqnarray}
 \delta\gamma = \gamma_{max}-\gamma_{min}=\frac{\pi}{2}-\gamma_{min}. 
 \label{E27}
\end{eqnarray}
A careful look at the stringy embedding further reveals that, $ \delta\gamma \sim \frac{|\Delta \varphi_{m} |}{|\Delta \varphi|}$. In order to arrive at the NR magnon dispersion relation one must therefore set the limit\footnote{This corresponds to solitonic excitations \emph{localized} along the equator of $  S^2$. }, $ |  \delta\gamma | \ll 1$ which upon substituting into (\ref{e25}) yields, $ \Delta\jmath \sim \delta\gamma \sim |\Delta\varphi|$ together with, $ |\Delta \varphi_{m} | \sim \Big|\frac{p_{m}}{c}\Big| \sim (\delta\gamma)^{2} $.

Putting all these facts together, we finally arrive at the cherished giant magnon dispersion relation of the following form,
\begin{eqnarray}
\mathcal{E} -\mathcal{J}_{\varphi}\sim \frac{\sqrt{\mathtt{g}}}{2\pi}|\Delta\varphi_{m}| \approx \frac{\sqrt{\mathtt{g}}}{2\pi} \Big|\frac{p_{m}}{c}\Big|
\label{e27}
\end{eqnarray}
which we interpret as single spin giant NR magnon in the dual SMT at strong coupling.

The other limiting condition we are interested in is the following,
\begin{eqnarray}
\gamma_{max}=\sin^{-1}\Big| \frac{\alpha}{\beta}\Big|
\end{eqnarray}
and, 
\begin{eqnarray}
\gamma_{min}=\sin^{-1}\Big| \frac{\xi}{\eta}\Big|
\end{eqnarray}
where, we set $ |\beta|\rightarrow \alpha $ which thereby yields, $ \gamma_{max}\sim \frac{\pi}{2} $. Substituting this limit into (\ref{e8}) one exactly recovers (\ref{e18}) and thereby the original NR magnon dispersion relation (\ref{e27}). This observation is related to the underlying fact that unlike the relativistic example \cite{Kruczenski:2004wg}-\cite{Ishizeki:2007we} both the conserved charges ($ \mathcal{E} $ and $ \mathcal{J}_{\varphi} $) as well as the deficit angle ($ \Delta\varphi $) are invariant under the exchange of the limits of the integral namely, $ \gamma_{max}\leftrightarrow \gamma_{min} $. The absence of single spike solutions \cite{Ishizeki:2007we} is what we identify as a special feature of NR string sigma models (over $ R\times S^2 $) in contrast to its relativistic cousins.

\subsection{Adding NS-NS fluxes}
\subsubsection{$1/c$ limit}
We now generalize our results in the presence of background fluxes \cite{Harmark:2019upf} and in particular confine our attention only to the NS-NS sector. Under these circumstances, the sigma model Lagrangian on $ R\times S^2 $  could therefore be formally expressed as, 
\begin{eqnarray}
\mathcal{S}_{NG}=\frac{\sqrt{\lambda}}{2\pi}\int d\tau d\sigma(\mathcal{L}_{NG}+\delta\mathcal{L}_{NG})
\label{e30}
\end{eqnarray}
where we identify the new NS-NS contribution \cite{Harmark:2019upf},
\begin{eqnarray}
\delta\mathcal{L}_{NG}=\frac{\varepsilon^{\alpha \alpha'}\varepsilon^{\beta \beta'}}{\varepsilon^{\alpha \alpha'}\chi_{\alpha}(\partial_{\alpha'}\zeta +\mathfrak{b}_{\alpha'})}(\mathfrak{b}_{\alpha'}\partial_{\beta'}\zeta +\mathfrak{b}_{\alpha'}\mathfrak{b}_{\beta'})\left( \partial_{\alpha}\theta \partial_{\beta}\theta +\sin^{2}\theta\partial_{\alpha}\varphi \partial_{\beta}\varphi\right)-\varepsilon^{\alpha \beta}\mathcal{B}_{\alpha \beta}
\end{eqnarray}
together with the following specification,
\begin{eqnarray}
\mathcal{B}_{\alpha \beta}=\partial_{\alpha}X^{M}\partial_{\beta}X^{N}\mathcal{B}_{MN}~;~\mathfrak{b}_{\alpha}=\partial_{\alpha}X^{M}\mathfrak{b}_{M}
\end{eqnarray}
where, $ X^{M} $s are the so called embedding coordinates\footnote{Notice that here $ X^{M}=\lbrace \zeta, X^{\mu}\rbrace $ where, $ \mu =\lbrace t, \theta , \varphi \rbrace$. Furthermore, we decompose the magnetic field as  \cite{Harmark:2019upf}, $ \mathcal{B}_{MN}\equiv \lbrace \mathcal{B}_{\mu \nu}, \mathcal{B}_{\zeta \mu}=\mathfrak{b}_{\mu} \rbrace $ and $ \mathfrak{b}_{M}=\lbrace \mathfrak{b}_{\mu}, \mathfrak{b}_{\zeta} \rbrace $.} along with $ \mathcal{B}_{\zeta \mu}=0 $, $  \mathfrak{b}_{\zeta}=1 $ \cite{Harmark:2019upf}.

Our next step would be to take $ 1/c $ limit of (\ref{e30}). The scaling limit corresponding to NS-NS fluxes are introduced as,
\begin{eqnarray}
 \mathfrak{b}_{\alpha}=c \partial_{\alpha}\tilde{\zeta}=c \tilde{\mathfrak{b}}_{\alpha}~;~\mathcal{B}_{\theta \varphi}=c \tilde{B}\sin \tilde{\theta}= c\tilde{\mathcal{B}}_{\theta \varphi}~;~ B= c \tilde{B}
 \label{e33}
\end{eqnarray}
together with \cite{Roychowdhury:2019olt},
\begin{eqnarray}
\chi_{\alpha}=c^2 \partial_{\alpha}\tilde{t}+\partial_{\alpha}\tilde{\psi}-\cos\tilde{\theta}\partial_{\alpha}\tilde{\varphi}~;~\mathfrak{h}_{\alpha \beta}=\partial_{\alpha}\theta \partial_{\beta}\theta +\sin^2\theta \partial_{\alpha}\varphi \partial_{\beta}\varphi \equiv \tilde{\mathfrak{h}}_{\alpha \beta}~;~\zeta = c \tilde{\zeta}.
\end{eqnarray}

Using (\ref{e5}) and (\ref{e33}), we finally obtain NS-NS contribution in the NR limit\footnote{As before we remove tildes from the polar angles.},
\begin{eqnarray}
\delta \tilde{S}_{NG}&=&-\frac{\sqrt{\mathtt{g}}\tilde{B}}{2\pi}\int d\tau d\sigma \varepsilon^{\alpha \beta}\partial_{\alpha}\theta \partial_{\beta}\varphi \sin\theta +\mathcal{O}(c^{-2})\nonumber\\
&\approx &\frac{\sqrt{\mathtt{g}}\tilde{B}\varpi}{2\pi}\int d\tau d\sigma \theta' \sin\theta \equiv -\frac{\sqrt{\mathtt{g}}}{2\pi}\int d\tau d\sigma \partial_{\sigma}(\tilde{B}\varpi \cos\theta)
\label{e35}
\end{eqnarray}
where we identify $ \tilde{B} $ as \emph{rescaled} background magnetic field \cite{Huang:2006vz}-\cite{Lee:2008sk} with fluxes attached to $ S^2 $. From (\ref{e35}), it is quite evident that the background NS-NS fluxes do not affect the equation of motion corresponding to $ \theta (\sigma)$ and hence for example, the poles appearing above in (\ref{E19}) also remain unchanged.
\subsubsection{Dispersion relation}
Below we note down conserved charges and their respective modifications due to the presence of background fluxes. To start with, notice that the energy (\ref{E19}) associated with the classical stringy configuration does not receive any correction due to the presence of NS-NS fluxes. Furthermore, the deficit angle (\ref{E18}) between the end points of the string remains unchanged too. Therefore the only change that appears to be in the dispersion relation (\ref{e27}) is through (\ref{E17}) namely,
\begin{eqnarray}
\mathcal{J}_{\varphi}\mapsto \mathcal{J}_{\varphi} +\Delta \mathcal{J}_{\varphi}
\end{eqnarray}
where we note down the constant shift in the angular momentum as,
\begin{eqnarray}
\Delta \mathcal{J}_{\varphi}\sim  \frac{\sqrt{\mathtt{g}}}{2\pi}\Big |\frac{B}{c} \Big | \cos\gamma_{min}.
\label{e38}
\end{eqnarray}

Finally, using (\ref{E27}) and (\ref{e38}) and considering the geometrical analogue of the NR limit ($ |\delta \gamma | \ll 1 $) we arrive at the giant magnon dispersion relation of the following form,
\begin{eqnarray}
\mathcal{E} -\mathcal{J}_{\varphi}\sim \frac{\sqrt{\mathtt{g}}}{2\pi}|\Delta\tilde{\varphi}_{m}| \approx \frac{\sqrt{\mathtt{g}}}{2\pi} \Big|\frac{\tilde{p}_{m}}{c}\Big|
\label{e39}
\end{eqnarray}
where we identify the R.H.S. of (\ref{e39}),
\begin{eqnarray}
\tilde{p}_{m}=p_{m}\left( 1+\Delta p_{m} \right) ~;~\Delta p_{m}\sim \Big |\frac{B}{c} \Big |\ll 1
\end{eqnarray}
as the \emph{effective} NR magnon momentum in the presence of background NS-NS fluxes.
\section{Summary and final remarks}
The conclusion as well as key observations of the present paper is as follows. We reestablish the strong coupling nonrelativistic (NR) giant magnon spectrum strating from NR rigidly rotating strings over torsional Newton-Cartan (TNC) geometry with $ R \times S^2 $ topology \cite{Grosvenor:2017dfs}. We further generalize our results considering background NS-NS fluxes and obtain effective NR magnon momentum in the presence of background magnetic field. All these findings strongly indicate towards NR sigma model/SMT correspondence \cite{Harmark:2017rpg}. However, quite surprisingly, we also notice that unlike its relativistic counterpart \cite{Ishizeki:2007we}, the NR sigma model lacks of single spike solutions. We identify this as a special feature associated to $ 1/c $ limit itself which may be of worth exploring further in the near future. \\ \\ 
{\bf {Acknowledgements :}}
 The author is indebted to the authorities of IIT Roorkee for their unconditional support towards researches in basic sciences. \\\\ 


\end{document}